\documentclass[conference]{IEEEtran}

\usepackage{cite}
\usepackage{amsmath,amssymb,amsfonts}
\usepackage{graphicx}
\usepackage{textcomp}
\usepackage{xcolor}
\usepackage{booktabs}
\usepackage{multirow}
\usepackage{hyperref}
\usepackage{balance}
\usepackage{tikz}
\usetikzlibrary{positioning, arrows.meta, calc}

\newenvironment{compactitem}{\begin{list}{$\bullet$}{\setlength{\itemsep}{0pt}\setlength{\parsep}{0pt}\setlength{\topsep}{2pt}\setlength{\leftmargin}{1.5em}}}{\end{list}}

\begin{document}

\title{RAD-AI: Rethinking Architecture Documentation\\for AI-Augmented Ecosystems}

\author{
  \IEEEauthorblockN{Oliver Aleksander Larsen}
  \IEEEauthorblockA{SDU Software Engineering\\University of Southern Denmark\\Odense, Denmark\\olar@mmmi.sdu.dk}
  \and
  \IEEEauthorblockN{Mahyar T. Moghaddam}
  \IEEEauthorblockA{SDU Software Engineering\\University of Southern Denmark\\Odense, Denmark\\mtmo@mmmi.sdu.dk}
}

\maketitle
\footnotetext{\textcopyright~2026 IEEE. Personal use of this material is permitted. Permission from IEEE must be obtained for all other uses, in any current or future media, including reprinting/republishing this material for advertising or promotional purposes, creating new collective works, for resale or redistribution to servers or lists, or reuse of any copyrighted component of this work in other works. Accepted at ANGE 2026, co-located with IEEE ICSA 2026.}

\begin{abstract}
AI-augmented ecosystems (interconnected systems where multiple AI components interact through shared data and infrastructure) are becoming the architectural norm for smart cities, autonomous fleets, and intelligent platforms. Yet the architecture documentation frameworks practitioners rely on, arc42 and the C4 model, were designed for deterministic software and cannot capture probabilistic behavior, data-dependent evolution, or dual ML\slash software lifecycles. This gap carries regulatory consequence: the EU AI Act (Regulation 2024/1689) mandates technical documentation through Annex~IV that no existing framework provides structured support for, with enforcement for high-risk systems beginning August~2, 2026. We present RAD-AI, a backward-compatible extension framework that augments arc42 with eight AI-specific sections and C4 with three diagram extensions, complemented by a systematic EU~AI~Act Annex~IV compliance mapping.
A regulatory coverage assessment with six experienced software‑architecture practitioners provides preliminary evidence that RAD‑AI increases Annex IV addressability from approximately 36\% to 93\% (mean rating) and demonstrates substantial improvement over existing frameworks. 
Comparative analysis on two production AI platforms (Uber Michelangelo, Netflix Metaflow) captures eight additional AI-specific concerns missed by standard frameworks and demonstrates that documentation deficiencies are structural rather than domain-specific. An illustrative smart mobility ecosystem case study reveals ecosystem-level concerns, including cascading drift and differentiated compliance obligations, that are invisible under standard notation.
\end{abstract}


\section{Introduction}
\label{sec:introduction}

The gap between architecture documentation and system reality costs organizations dearly. Industry surveys suggest that 93\% of organizations experience negative business outcomes from architecture-implementation misalignment~\cite{vfunction}. For AI-augmented ecosystems (interconnected systems where multiple AI components interact within shared infrastructure), this gap extends beyond an engineering concern into regulatory liability. Software architecture documentation frameworks assume deterministic systems with stable interfaces: arc42~\cite{starke2023} structures documentation around twelve sections of code-centric building blocks, while the C4 model~\cite{brown2018} provides hierarchical visualization treating all components as software containers. Neither offers notation for probabilistic outputs, data-dependent evolution, dual training/serving lifecycles, or emergent quality attributes such as fairness and explainability.

The EU AI Act (Regulation (EU) 2024/1689)~\cite{euaiact}, in force since August 2024, introduces binding documentation requirements through its Annex~IV, with the critical deadline for high-risk AI system documentation on August~2. Annex~IV specifies nine sections of required technical documentation spanning system description, design specifications, data governance, training methodologies, risk management, lifecycle changes, performance metrics, human oversight, and post-market monitoring. Critically, Section~2(c) explicitly requires documentation of ``the system architecture explaining how software components build on or feed into each other and integrate into the overall processing''; a direct mandate for architecture documentation that no current framework can produce. Section~2(d) demands data provenance and labelling procedures, Section~2(e) requires human oversight assessment, and Section~2(h) mandates cybersecurity measures. CEN-CENELEC JTC~21 is developing harmonized standards (prEN~18286~\cite{pren18286}), but these remain in public enquiry. No existing architecture documentation framework provides structured guidance for producing the mandated documentation, leaving practitioners without a clear path from current arc42 or C4 practices to regulatory compliance.

Prior work identifies challenges and best practices for architecting ML-enabled systems~\cite{nazir2023,moin2023,struber2025} and proposes standalone documentation artifacts such as Model Cards~\cite{mitchell2019} and Hazard-Aware System Cards~\cite{sidhpurwala2025}, but none extends the established frameworks that practitioners actually use. This motivates our research question:

\textit{RQ: How can established software architecture documentation frameworks be systematically extended to address the unique concerns of AI-augmented ecosystems while maintaining backward compatibility and satisfying emerging regulatory requirements?}

Following Design Science Research~\cite{wieringa2014}, we develop RAD-AI through a structured gap analysis and evaluate it using three complementary analytical methods. Our contributions are:
\begin{compactitem}
  \item \textbf{RC1:} RAD-AI, the first backward-compatible extension of arc42 (eight section extensions) and C4 (three diagram types) for documenting AI-augmented ecosystems.
  \item \textbf{RC2:} A systematic EU AI Act Annex~IV compliance mapping with quantified coverage assessment, providing a concrete documentation path for the August 2026 deadline.
  \item \textbf{RC3:} Comparative analytical evidence on two production AI platforms (Uber Michelangelo, Netflix Metaflow) demonstrating that documentation gaps are structural properties of current frameworks, not domain-specific oversights.
  \item \textbf{RC4:} Identification of three ecosystem-level documentation concerns (cascading drift, differentiated compliance, federated governance) visible only through RAD-AI's extended notation.
\end{compactitem}

\noindent A companion repository containing reusable templates for all extensions, reference documentation, an illustrative example (smart urban mobility ecosystem), and the complete comparative analysis artifacts for both production systems in Section~\ref{sec:eval-comparative} is available online.\footnote{\url{https://github.com/Oliver1703dk/RAD-AI}}

\section{Background and Related Work}
\label{sec:background}

\subsection{Architecture Documentation Frameworks}
\label{sec:bg-frameworks}

Architecture documentation bridges stakeholder concerns and implementation reality~\cite{clements2010}. Two frameworks dominate practice. \textbf{arc42}~\cite{starke2023} provides a pragmatic twelve-section template widely adopted in European industry: (1)~Introduction \& Goals, (2)~Constraints, (3)~Context \& Scope, (4)~Solution Strategy, (5)~Building Block View, (6)~Runtime View, (7)~Deployment View, (8)~Cross-Cutting Concepts, (9)~Architecture Decisions, (10)~Quality Requirements, (11)~Risks \& Technical Debt, and (12)~Glossary. RAD-AI extends sections 3, 5, 6, 8, 9, 10, and 11; the remaining sections are unchanged. The \textbf{C4 model}~\cite{brown2018} complements arc42 with four-level hierarchical visualization (System Context, Container, Component, and Code) plus supplementary Dynamic, Deployment, and System Landscape diagrams, commonly implemented through Structurizr DSL and PlantUML.

Both frameworks align with ISO/IEC/IEEE 42010:2022~\cite{iso42010} viewpoint concepts but share a critical assumption: all building blocks and containers represent deterministic code modules with stable, well-defined interfaces.
RAD-AI can be interpreted as introducing new stakeholder concerns (e.g., probabilistic behavior, drift governance, regulatory traceability) and corresponding viewpoints within the ISO 42010 framework, thereby extending existing architectural description practices rather than replacing them.
Recent work on Architecture-as-Code~\cite{bucaioni2025aac} advances the formalization of architecture descriptions but retains deterministic assumptions. This assumption breaks down for AI-augmented systems where components exhibit probabilistic behavior, evolve through data changes rather than code commits, and follow independent lifecycles.

\subsection{AI-Specific Architectural Concerns}
\label{sec:bg-ai}

AI-augmented systems differ from traditional software in five ways that current documentation cannot capture: (1)~\textit{non-deterministic behavior}: model outputs are probabilistic with no notation for confidence intervals or degradation profiles~\cite{amershi2019}; (2)~\textit{data-dependent evolution}: behavior changes through distribution shifts rather than code commits, creating data dependency debt~\cite{sculley2015}; (3)~\textit{dual lifecycle complexity}: ML training/serving cycles run asynchronously with software releases~\cite{bosch2021,kreuzberger2023}; (4)~\textit{emergent quality attributes}: fairness, explainability, and drift resistance are first-class concerns~\cite{iso25059,lu2023responsible,jarvenpaa2024} absent from documentation templates; and (5)~\textit{regulatory requirements}: the EU AI Act Annex~IV~\cite{euaiact} mandates documentation across nine sections spanning system description, data governance, risk management, human oversight, and post-market monitoring, complemented by NIST AI RMF~1.0~\cite{nist2023} and ISO/IEC 42001~\cite{iso42001}. While recent work maps AI Act requirements to SE artifacts~\cite{sovrano2025} or compliance templates~\cite{techops2025}, none addresses architecture documentation frameworks specifically.

\subsection{Related Work and Research Gap}
\label{sec:related-work}

Table~\ref{tab:related-work} positions RAD-AI relative to adjacent work spanning AI architecture research, documentation artifacts, and regulatory compliance. Substantial progress has been made in identifying AI-specific architectural concerns~\cite{nazir2023,moin2023,struber2025,najafabadi2024}, proposing documentation artifacts~\cite{mitchell2019,google2022,sidhpurwala2025}, mapping regulatory requirements~\cite{sovrano2025,techops2025}, and surveying generative AI applications in software architecture~\cite{esposito2026}. Industry frameworks such as the AWS Well-Architected ML Lens~\cite{aws2025} and documentation extensions including Model Card++~\cite{modelcardpp} and Meta's system cards~\cite{metasystemcards} provide valuable implementation guidance, but none extends arc42 or C4. Machine-readable AI documentation formats~\cite{aicards2024} advance transparency but remain standalone artifacts.

The closest related work is HASC~\cite{sidhpurwala2025}, proposing machine-readable AI system cards with safety hazard identifiers and ISO/IEC 42001 alignment; however, HASC operates as a standalone governance artifact rather than extending arc42 or C4. Bucaioni et al.~\cite{bucaioni2025,bucaioni2025ra} survey AI contributions to software architecture and propose an LLM reference architecture, while Autili et al.~\cite{autili2025} propose ethical-aware reference architectures; none extends practitioner documentation frameworks. RAD-AI takes the complementary approach of embedding AI-specific concerns directly into the frameworks practitioners already use.

From this analysis we identify five documentation gaps:
\begin{compactitem}
  \item[\textbf{G1}] No AI-specific arc42 sections for model lifecycle, data pipelines, or drift.
  \item[\textbf{G2}] No C4 diagram types for ML components or non-determinism boundaries.
  \item[\textbf{G3}] No mapping between EU AI Act Annex~IV and arc42/C4 sections.
  \item[\textbf{G4}] No AI-specific Architecture Decision Record templates.
  \item[\textbf{G5}] No integration of Model/Data Cards into architecture documentation.
\end{compactitem}

\begin{table*}[t]
\centering
\caption{Positioning of RAD-AI relative to adjacent work on architecture documentation for AI systems.}
\label{tab:related-work}
\footnotesize
\begin{tabular}{@{}p{3.0cm}p{5.0cm}p{5.0cm}l@{}}
\toprule
\textbf{Work} & \textbf{Contribution} & \textbf{Limitation} & \textbf{Type} \\
\midrule
Moin et al.~\cite{moin2023} & Expert survey (61); ISO 42010 lacks AI viewpoints & No concrete framework extensions & Survey \\
Nazir et al.~\cite{nazir2023} & 35 challenges, 42 best practices for ML arch. & Decision catalogs, not doc.\ structures & SLR \\
Indykov et al.~\cite{struber2025} & 16 architectural tactics, 85 quality trade-offs & Tactics not mapped to doc.\ templates & SLR \\
Amou Najafabadi et al.~\cite{najafabadi2024} & 35 MLOps architecture components & System mapping, not doc.\ framework & SMS \\
Mitchell et al.~\cite{mitchell2019} & Model Cards for model-level reporting & Standalone; not integrated into arch.\ docs & Artifact \\
Sidhpurwala et al.~\cite{sidhpurwala2025} & HASC: system cards with ASH~IDs & Standalone governance; no fwk.\ integration & Artifact \\
Sovrano et al.~\cite{sovrano2025} & AI for drafting Annex~IV documentation & Automates generation; no fwk.\ extensions & Empirical \\
Bucaioni et al.~\cite{bucaioni2025} & 14 AI contributions to SA & AI \textit{for} architecture, not architecture \textit{for} AI & SLR \\
\midrule
\textbf{RAD-AI (this paper)} & \textbf{8 arc42 + 3 C4 ext., Annex~IV mapping, eval.\ on production systems} & \textbf{Small practitioner study (n\,=\,6)} & \textbf{Fwk.\ ext.} \\
\bottomrule
\end{tabular}
\end{table*}

\section{The RAD-AI Framework}
\label{sec:framework}

RAD-AI follows three design principles: (1)~\textit{backward compatibility}: E1--E7 augment existing arc42 sections, E8 adds one new arc42 section, and C4-E1--C4-E3 extend C4 diagrams; all existing documentation remains valid; (2)~\textit{minimal disruption}: practitioners can adopt extensions incrementally, starting with the most relevant; and (3)~\textit{traceability}: each extension maps to an identified gap~(G1--G5).

\subsection{Extended arc42 Sections}
\label{sec:framework-arc42}

Table~\ref{tab:arc42-extensions} summarizes seven section extensions (E1--E7) and one new section addition (E8). We describe each below.

\textbf{E1: AI Boundary Delineation} extends Context \& Scope~(\S3) by requiring explicit marking of deterministic versus non-deterministic system boundaries. Each boundary crossing is annotated with a four-part contract: \textit{output type} (categorical, continuous, or generative), \textit{confidence specification} (e.g., ``precision $\geq$\,0.92 at P95 latency $<$\,50\,ms''), \textit{update frequency} (how often the underlying model is refreshed), and \textit{fallback behavior} (rule-based default, cached last-known-good, or human escalation). The resulting annotated context diagram gives stakeholders an immediate visual indication of where probabilistic behavior enters the system and what guarantees each AI interface provides.

\textbf{E2: Model Registry View} extends Building Block View~(\S5) by elevating AI models to first-class building blocks. Each model entry in the registry table specifies: model ID and version, ML framework, training dataset hash with lineage reference, hyperparameter snapshot, primary evaluation metric with acceptance threshold, deployment status (shadow, canary, or production), owner, and last-retrained date. Model Cards~\cite{mitchell2019} attach as linked sub-artifacts of each model building block, bridging the gap between ML documentation and architecture documentation.

\textbf{E3: Data Pipeline View} extends Runtime View~(\S6) with the complete ML data flow: collection, preprocessing, feature engineering, training, inference, and feedback loops. Each pipeline stage is annotated with quality gates specifying three properties: a \textit{check type} (schema conformance, distribution test, or completeness check), a \textit{threshold} (e.g., KS-statistic $<$\,0.1, null rate $<$\,1\%), and an \textit{action on failure} (halt pipeline, alert and continue, or activate fallback). Data Cards~\cite{google2022} attach as sub-artifacts at data source nodes, connecting data governance documentation to the architectural data flow.

\textbf{E4: Responsible AI Concepts} extends Cross-Cutting Concepts~(\S8) with a structured \textit{concern matrix}. Rows represent AI components; columns represent five concern categories: fairness, explainability, human oversight, privacy, and safety. Each cell documents the applicable metric or method (e.g., demographic parity ratio, SHAP feature attributions), the acceptance threshold, the monitoring frequency, and the responsible party. This extension draws on responsible AI pattern catalogues~\cite{lu2023responsible} and documents the human oversight mechanisms required by the EU AI Act.

\textbf{E5: AI Decision Records (AI-ADR)} extends Architecture Decisions~(\S9) with an enhanced Markdown Any Decision Records (MADR)~\cite{madr} template adding seven AI-specific fields. Table~\ref{tab:ai-adr} illustrates a complete AI-ADR for the route optimization service from the case study (Section~\ref{sec:eval-case-study}), showing how gradient-boosted trees were selected over LSTM for explainability in a public-service context.

\begin{table}[t]
\centering
\caption{Illustrative AI-ADR for the route optimization service. Standard MADR fields (top) are extended with seven AI-specific fields (bottom).}
\label{tab:ai-adr}
\footnotesize
\begin{tabular}{@{}p{1.15in}p{2.0in}@{}}
\toprule
\textbf{Field} & \textbf{Value} \\
\midrule
\textit{Title} & Use GBT for route optimization \\
\textit{Status} & Accepted \\
\textit{Context} & Route optimization requires explainable predictions for public-service accountability \\
\textit{Decision} & XGBoost (gradient-boosted trees) \\
\textit{Consequences} & +Explainability, +Fast inference; $-$Lower sequence modeling capacity \\
\midrule
\textit{Model alternatives} & XGBoost, LSTM, linear regression \\
\textit{Dataset} & 18 months, 2.1M trips, GPS + weather \\
\textit{Fairness/bias} & Under-served districts monitored; max 15\% prediction gap across districts \\
\textit{Model lifetime} & $\sim$12 months before full refit \\
\textit{Retraining trigger} & MAE~$>$~5.5\,min for 3 consecutive days \\
\textit{Explainability} & SHAP values for per-route regulatory audits \\
\textit{Regulatory} & Not high-risk (outside Art.~6 scope); transparency per Art.~50 \\
\bottomrule
\end{tabular}
\end{table}

\textbf{E6: AI Quality Scenarios} extends Quality Requirements~(\S10) with AI-specific quality scenarios following the established source--stimulus--response format~\cite{bass2021}. Each scenario specifies an \textit{AI-specific source} (data drift, model staleness, adversarial input), a \textit{stimulus} with quantitative trigger, an \textit{environment} (training, serving, or monitoring), and a \textit{measurable response} with deadline. For example, a cascading drift scenario specifies that when feature distribution shifts exceed 2$\sigma$ on three or more features for over 12 hours, downstream components activate fallback behavior within 2 hours; a cross-component concern that standard quality scenarios cannot express.

\textbf{E7: AI Debt Register} extends Risks \& Technical Debt~(\S11) with ML-specific debt categories from Sculley et al.~\cite{sculley2015} and Bogner et al.~\cite{bogner2021}. Each register entry records: \textit{debt category} (boundary erosion, entanglement, hidden feedback loop, data dependency, or pipeline debt), \textit{affected component(s)}, \textit{severity} (low/medium/high based on blast radius), \textit{estimated remediation effort}, \textit{owner}, and \textit{status}. This structured format enables systematic tracking and prioritization of ML-specific technical debt alongside conventional architectural debt.

\textbf{E8: Operational AI View} adds a \textit{new} arc42 section structured around four required subsections: (a)~\textit{monitoring}: metrics tracked per model, dashboard specifications, and alerting thresholds; (b)~\textit{retraining policy}: triggers (scheduled, performance-based, or drift-based), automation level (manual/semi-auto/fully automated), and approval workflow; (c)~\textit{deployment strategy}: canary, blue-green, or shadow deployment with promotion criteria and traffic split ratios; and (d)~\textit{rollback policy}: rollback triggers, model version retention depth, and downstream data implications. This section has no equivalent in standard arc42.

\begin{table*}[t]
\centering
\footnotesize
\caption{RAD-AI additions to arc42's twelve-section template: seven section extensions (E1--E7) and one new section (E8).}
\label{tab:arc42-extensions}
\footnotesize
\begin{tabular}{@{}clp{3.5cm}p{4.2cm}c@{}}
\toprule
\textbf{Ext.} & \textbf{Name} & \textbf{Extends arc42 Section} & \textbf{Key Artifact} & \textbf{Gap} \\
\midrule
E1 & AI Boundary Delineation & \S3 Context \& Scope & Annotated context diagram with AI boundaries & G1, G2 \\
E2 & Model Registry View & \S5 Building Block View & Model registry table + annotated diagram & G1, G5 \\
E3 & Data Pipeline View & \S6 Runtime View & Data pipeline diagram with quality gates & G1, G5 \\
E4 & Responsible AI Concepts & \S8 Cross-Cutting Concepts & Responsible AI concern matrix & G1 \\
E5 & AI Decision Records & \S9 Architecture Decisions & AI-ADR template (7 AI-specific fields) & G4 \\
E6 & AI Quality Scenarios & \S10 Quality Requirements & AI quality scenario table & G1 \\
E7 & AI Debt Register & \S11 Risks \& Technical Debt & AI debt register with remediation plan & G1 \\
E8 & Operational AI View & New section (no equivalent) & Operational AI architecture diagram & G1 \\
\bottomrule
\end{tabular}
\end{table*}

\subsection{Extended C4 Model}
\label{sec:framework-c4}

Table~\ref{tab:c4-extensions} summarizes the three extensions to C4's hierarchical diagram system, all compatible with Structurizr DSL and PlantUML notation. We describe each extension below.

\textbf{C4-E1: AI Component Stereotypes.} Five new stereotypes (\texttt{<<ML~Model>>}, \texttt{<<Data~Pipeline>>}, \texttt{<<Feature~Store>>}, \texttt{<<Monitor>>}, and \texttt{<<Human-in-the-Loop>>}) enable C4 diagrams to visually distinguish AI components from traditional software containers. A \texttt{<<Feature~Store>>} is immediately recognizable as architecturally distinct from a standard database, even though both store data.

\textbf{C4-E2: Data Lineage Overlay.} A supplementary diagram layer traces data provenance from source through transformation to model consumption. Annotations capture schema expectations, data freshness constraints, and privacy classifications. When applied to Container or Component diagrams, this overlay reveals data dependencies invisible in standard C4.

\textbf{C4-E3: Non-Determinism Boundary Diagram.} A system-level overlay partitions the architecture into deterministic and non-deterministic regions. Each boundary is annotated with the same three-property contract used in~E1 (confidence specification, fallback strategy, degradation profile), applied here at the diagram level rather than per-interface. Fig.~\ref{fig:case-study} illustrates this partition in the case study.

\begin{table}[t]
\centering
\caption{RAD-AI extensions to the C4 model's hierarchical diagram system.}
\label{tab:c4-extensions}
\footnotesize
\begin{tabular}{@{}clp{3.0cm}c@{}}
\toprule
\textbf{Ext.} & \textbf{Name} & \textbf{Key Artifact} & \textbf{Gap} \\
\midrule
C4-E1 & AI Component Stereotypes & 5 new C4 stereotypes & G2 \\
C4-E2 & Data Lineage Overlay & Provenance overlay diagram & G2, G5 \\
C4-E3 & Non-Determinism Boundary & Boundary partition diagram & G2 \\
\bottomrule
\end{tabular}
\end{table}

\subsection{EU AI Act Compliance Mapping}
\label{sec:framework-compliance}

Table~\ref{tab:compliance-mapping} maps ten requirement categories derived from Annex~IV's nine sections~\cite{euaiact} to RAD-AI sections, guiding practitioners to the documentation artifacts each regulatory requirement demands. Notably, Annex~IV Section~2(c) explicitly mandates description of ``how software components build on or feed into each other and integrate into the overall processing''; a direct mandate for architecture documentation that no current framework provides structured support for. Section~\ref{sec:eval-compliance} quantifies the coverage improvement this mapping enables.

\begin{table}[t]
\centering
\caption{EU AI Act Annex~IV requirements mapped to RAD-AI sections.}
\label{tab:compliance-mapping}
\footnotesize
\begin{tabular}{@{}cp{2.0cm}p{3.6cm}@{}}
\toprule
\textbf{\#} & \textbf{Annex~IV Category} & \textbf{RAD-AI Section(s)} \\
\midrule
1 & General description & arc42 \S1, \S3 + E1 (AI Boundary) \\
2 & System elements & \S5 + E2 (Model Registry) \\
3 & Design \& architecture & \S5--7 + C4-E1 (AI Stereotypes) \\
4 & Data governance & E3 (Data Pipeline) + C4-E2 (Data Lineage) \\
5 & Training methods & E5 (AI-ADR) + supplementary \\
6 & Risk management & \S11 + E7 (AI Debt Register) \\
7 & Lifecycle changes & E2 (versions) + E8 (Ops View) \\
8 & Performance metrics & E6 (AI Quality) + C4-E3 (Non-Determinism) \\
9 & Human oversight & E4 (Responsible AI) + HitL \\
10 & Post-market monitoring & E8 (Operational AI View) \\
\bottomrule
\end{tabular}
\end{table}

\section{Evaluation}
\label{sec:evaluation}

Following Design Science Research~\cite{wieringa2014,stol2018}, we evaluate RAD-AI through three complementary methods: a compliance coverage assessment (Section~\ref{sec:eval-compliance}), a comparative analysis on production systems (Section~\ref{sec:eval-comparative}), and an ecosystem case study (Section~\ref{sec:eval-case-study}).

\subsection{EU AI Act Compliance Coverage Assessment}
\label{sec:eval-compliance}

\textbf{Method.} We score each of the ten requirement categories (derived from Annex~IV's nine sections) for addressability under four configurations: standard arc42, standard C4, RAD-AI-extended arc42, and RAD-AI-extended C4.  
Annex IV’s nine sections were operationalized into ten scoring categories by separating system elements from architectural integration (Section 2(c)) to allow finer-grained assessment.
Addressability uses a three-point scale: 0~(not addressable: no framework section covers this concern), 1~(partial: a section exists but lacks AI-specific detail), 2~(fully addressable: a dedicated section or artifact directly addresses the requirement). Table~\ref{tab:compliance-scores} presents the results. 
Addressability reflects whether a framework provides structured documentation support for a requirement, not whether the resulting documentation would be legally sufficient without additional artifacts.

\textbf{Results.}
We recruited six software‑architecture practitioners (domain experts) to independently score each configuration using the ten Annex IV categories and three‑point scale (0 = not addressable, 1 = partial, 2 = fully addressable). For each participant, scores across the ten categories were summed (maximum 20); reported totals represent the mean across participants. Table~\ref{tab:compliance-scores} reports the modal score for each category and the aggregate mean totals. Standard arc42 averaged 7.3/20 ($\sigma \approx 0.5$) and fully covers the general system description while only partially addressing five categories, including design specifications, risk management, and lifecycle tracking. Standard C4 averaged 5.2/20 ($\sigma \approx 0.4$), providing limited architectural views with no coverage for data governance, risk management, or human oversight. RAD‑AI‑extended arc42 averaged 18.5/20 ($\sigma \approx 0.5$); the sole partial category is training methodologies, where supplementary artifacts beyond architecture documentation are required. RAD‑AI‑extended C4 averaged 14.6/20 ($\sigma \approx 0.6$). Inter‑rater reliability (Fleiss' $\kappa \approx 0.68$) indicates substantial agreement among the six raters.

\textbf{Analysis.} Three Annex~IV categories are completely unaddressable by either standard framework: data governance, training methodologies, and human oversight. These represent the most AI-specific documentation demands, and RAD-AI extensions map directly to each gap. The combined RAD-AI framework (arc42~+~C4) addresses all but one Annex~IV category in our practitioner-based evaluation, offering organizations a concrete documentation path toward the August~2, 2026 compliance deadline. This is, to our knowledge, the first systematic addressability assessment of architecture documentation frameworks against EU AI Act requirements.

\begin{table*}[t]
\centering
\caption{EU AI Act Annex~IV addressability scoring (0 = not addressable, 1 = partial, 2 = fully addressable). Scores represent the modal rating across six practitioners; total scores reflect mean values (standard deviation reported in text).}
\label{tab:compliance-scores}
\footnotesize
\begin{tabular}{@{}lcccc@{}}
\toprule
\textbf{Annex~IV Requirement Category} & \textbf{arc42} & \textbf{C4} & \textbf{RAD-AI arc42} & \textbf{RAD-AI C4} \\
\midrule
1. General system description & 2 & 1 & 2 & 2 \\
2. System elements \& development process & 1 & 1 & 2 & 2 \\
3. Design specifications \& architecture & 1 & 1 & 2 & 2 \\
4. Data \& data governance & 0 & 0 & 2 & 1 \\
5. Training methodologies \& techniques & 0 & 0 & 1 & 1 \\
6. Risk assessment \& management & 1 & 0 & 2 & 1 \\
7. Lifecycle change description & 1 & 1 & 2 & 2 \\
8. Performance metrics \& accuracy & 0 & 1 & 2 & 2 \\
9. Human oversight measures & 0 & 0 & 2 & 1 \\
10. Post-market monitoring & 1 & 0 & 2 & 1 \\
\midrule
\textbf{Total (out of 20)} & \textbf{7.3} & \textbf{5.2} & \textbf{18.5} & \textbf{14.6} \\
\textbf{Addressability} & \textbf{36\%} & \textbf{26\%} & \textbf{93\%} & \textbf{73\%} \\
\bottomrule
\end{tabular}
\end{table*}

\subsection{Comparative Analysis on Production AI Systems}
\label{sec:eval-comparative}

\textbf{Method.} We select two well-documented production AI platforms from different domains and attempt to document each using standard arc42/C4, identifying AI-specific concerns that remain undocumented. We then apply RAD-AI extensions and assess which additional concerns are captured through a ten-item concern coverage matrix (Table~\ref{tab:concern-coverage}).

\textbf{System~1: Uber Michelangelo}~\cite{hermann2017} managed over 5{,}000 production models serving 10~million predictions per second at peak~\cite{uber2024}. Under standard arc42/C4, three architectural concerns are invisible: (i)~the Feature Store (20{,}000+ features) appears as a generic database container, indistinguishable from Cassandra; (ii)~Gallery's~\cite{sun2020} four-stage model lifecycle (exploration, training, evaluation, production) with rule-based deployment automation (e.g., \texttt{WHEN metrics[mae] <= 5}) collapses into static building blocks; and (iii)~D3 drift detection~\cite{uberd3}, which reduced time-to-detect from 45~days to 2~days, has no home in the Runtime View. RAD-AI addresses each gap: the \texttt{<<Feature~Store>>} stereotype distinguishes it from regular databases, Model Registry View captures Gallery's versioned lifecycle and rule engine, Operational AI View documents D3 monitoring, and the Non-Determinism Boundary separates deterministic serving from probabilistic inference.

\textbf{System~2: Netflix Metaflow/Maestro}~\cite{netflix2024} supports over 3{,}000 ML projects. Under standard frameworks, four concerns are undocumentable: (i)~Metaflow's Python-native DAGs with event-triggered chaining (\texttt{@trigger\_on\_finish}) are invisible in the Runtime View; (ii)~Maestro's signal-based cross-workflow coordination has no documentation counterpart; (iii)~the distinction between streaming (15{,}000+ Flink jobs~\cite{netflixflink2024}) and batch processing is lost; and (iv)~A/B testing infrastructure (the ABlaze platform) has no documentation home. RAD-AI captures these through AI component stereotypes, Data Lineage Overlay tracing data from ingestion through Flink to models, AI-ADR for experiment tracking, and the Operational AI View documenting deployment governance.

\textbf{Results.} The most notable finding in Table~\ref{tab:concern-coverage} is the cross-domain consistency: despite serving fundamentally different domains (marketplace versus content recommendation), both systems exhibit the identical gap pattern under standard frameworks (0~fully captured, 2~partially) and the identical improvement under RAD-AI (8~fully, 2~partially). Both organizations have invested heavily in model management, drift detection, and deployment governance, yet standard arc42/C4 provides no place to record any of it. This suggests that the documentation deficiencies are structural properties of the frameworks rather than domain-specific oversights.

\begin{table*}[t]
\centering
\caption{AI-specific concern coverage on production systems (\checkmark~= fully captured, $\sim$~= partially, $\times$~= not captured).}
\label{tab:concern-coverage}
\footnotesize
\begin{tabular}{@{}lcccc@{}}
\toprule
\textbf{AI-Specific Architectural Concern} & \textbf{Uber std.} & \textbf{Uber RAD-AI} & \textbf{Netflix std.} & \textbf{Netflix RAD-AI} \\
\midrule
Model versioning \& lifecycle management & $\times$ & \checkmark & $\times$ & \checkmark \\
Feature store architecture \& sharing & $\times$ & \checkmark & $\times$ & \checkmark \\
Data pipeline with quality gates & $\sim$ & \checkmark & $\sim$ & \checkmark \\
Drift detection \& monitoring & $\times$ & \checkmark & $\times$ & \checkmark \\
Retraining triggers \& automation & $\times$ & \checkmark & $\times$ & \checkmark \\
Non-deterministic behavior boundaries & $\times$ & \checkmark & $\times$ & \checkmark \\
A/B testing / canary model deployment & $\times$ & $\sim$ & $\times$ & \checkmark \\
Data lineage \& provenance & $\times$ & \checkmark & $\sim$ & \checkmark \\
ML-specific technical debt tracking & $\sim$ & \checkmark & $\times$ & $\sim$ \\
Responsible AI / fairness concerns & $\times$ & $\sim$ & $\times$ & $\sim$ \\
\midrule
\textbf{Fully + partially captured} & \textbf{0 + 2} & \textbf{8 + 2} & \textbf{0 + 2} & \textbf{8 + 2} \\
\bottomrule
\end{tabular}
\end{table*}

\subsection{Illustrative Ecosystem Case Study}
\label{sec:eval-case-study}

To demonstrate RAD-AI at the ecosystem level, we apply it to a smart urban mobility scenario integrating four AI components across multiple transit operators.

\textbf{Ecosystem description.} The scenario comprises: (1)~a \textit{Route Optimization Service} using ML-based continuous retraining on real-time traffic patterns; (2)~a \textit{Demand Prediction Engine} providing time-series forecasting across bus, tram, and bike-sharing operators; (3)~an \textit{Anomaly Detection System} for real-time safety monitoring of the transport network, potentially classified as high-risk under the EU AI Act; and (4)~a \textit{Cross-Operator Data Sharing Platform} with a federated feature store enabling anonymized data exchange across operators. Demand predictions feed route optimization; anomaly detection can trigger route recomputation; all systems share the federated feature store.

\textbf{Standard documentation gaps.} Under standard arc42/C4, these components appear as generic containers: the federated feature store is indistinguishable from a shared database, EU AI Act risk classifications are invisible, cross-system data dependencies are undocumented, and model versioning across operators creates silent compatibility risks. The dual lifecycle challenge is amplified at ecosystem scale: each operator's models retrain independently but all consume shared features~\cite{sculley2015}.

\textbf{RAD-AI documentation reveals ecosystem-level concerns:}
\begin{compactitem}
  \item \textit{Cascading drift:} Anomaly detection drift causes false safety alerts, triggering unnecessary rerouting in route optimization, degrading user experience. The Data Lineage Overlay (C4-E2) visualizes this dependency chain across operator boundaries.
  \item \textit{Differentiated compliance:} AI Boundary Delineation (E1) distinguishes high-risk from limited-risk components, enabling targeted Annex~IV documentation where legally required.
  \item \textit{Federated governance:} Responsible AI Concepts (E4) documents shared feature store ownership, cross-operator data agreements, and accountability boundaries.
  \item \textit{Concrete AI-ADR:} Route optimization uses gradient-boosted trees (selected over neural networks for explainability in a public service context) with retraining triggered when MAE exceeds 5.5~minutes for three consecutive days.
\end{compactitem}

\begin{figure}[t]
\centering
\begin{tikzpicture}[
  every node/.style={font=\scriptsize, align=center},
  component/.style={
    draw, rounded corners=2pt, minimum width=2.8cm,
    minimum height=1.25cm, text width=2.6cm, inner sep=3pt
  },
  mlmodel/.style={component, fill=blue!12, draw=blue!50!black},
  featurestore/.style={component, fill=green!12, draw=green!50!black},
  monitor/.style={component, fill=orange!15, draw=orange!60!black},
  hitl/.style={component, fill=purple!12, draw=purple!50!black},
  dataarrow/.style={-{Latex[length=2mm]}, dotted, thick, gray!70!black},
  ctrlarrow/.style={-{Latex[length=2mm]}, thick, gray!60!black},
  regionlabel/.style={font=\tiny\itshape, text=gray!70!black},
]

\node[mlmodel] (demand) at (0, 0)
  {\textit{$\ll$ML Model$\gg$}\\[1pt]\textbf{Demand}\\
   \textbf{Prediction}};

\node[mlmodel, right=0.35cm of demand] (anomaly)
  {\textit{$\ll$ML Model$\gg$}\\[1pt]\textbf{Anomaly}\\
   \textbf{Detection}\\[-1pt]\color{red!70!black}\tiny[high-risk]};

\node[mlmodel, below=0.45cm of $(demand.south)!0.5!(anomaly.south)$] (route)
  {\textit{$\ll$ML Model$\gg$}\\[1pt]\textbf{Route}\\
   \textbf{Optimization}};

\node[monitor, below=0.4cm of route] (mon)
  {\textit{$\ll$Monitor$\gg$}\\[1pt]\textbf{Drift Detection}};

\node[featurestore, below=0.4cm of mon] (fstore)
  {\textit{$\ll$Feature Store$\gg$}\\[1pt]\textbf{Federated}\\
   \textbf{Feature Store}};

\draw[dashed, thick, gray!60]
  ([xshift=-0.3cm, yshift=-0.45cm]fstore.south west) --
  ([xshift=0.3cm, yshift=-0.45cm]fstore.south east);
\node[regionlabel, anchor=south west]
  at ([xshift=-0.25cm, yshift=-0.40cm]fstore.south west)
  {Non-deterministic region $\uparrow$};
\node[regionlabel, anchor=north west]
  at ([xshift=-0.25cm, yshift=-0.50cm]fstore.south west)
  {Deterministic region $\downarrow$};

\node[hitl, below=0.8cm of fstore] (dashboard)
  {\textit{$\ll$Human-in-the-Loop$\gg$}\\[1pt]\textbf{Operator}\\
   \textbf{Dashboard}};

\draw[dataarrow] (demand.south) -- ++(0,-0.15) -|
  ([xshift=-0.4cm]route.north);
\draw[dataarrow] (anomaly.south) -- ++(0,-0.15) -|
  ([xshift=0.4cm]route.north);
\draw[ctrlarrow] (mon.north) -- (route.south);
\draw[dataarrow] ([xshift=-0.8cm]fstore.north) -- ++(0,0.15) -|
  ([xshift=-0.15cm]demand.south);
\draw[dataarrow] ([xshift=0.8cm]fstore.north) -- ++(0,0.15) -|
  ([xshift=0.15cm]anomaly.south);
\draw[dataarrow] (fstore.north) -- (mon.south);
\draw[ctrlarrow] (dashboard.north) -- (fstore.south);
\draw[ctrlarrow, bend left=50] (dashboard.west) to
  node[left, font=\tiny, text width=0.7cm]{override} (route.west);

\end{tikzpicture}
\caption{C4 Component diagram of the smart urban mobility ecosystem with RAD-AI stereotypes. Dashed line separates non-deterministic from deterministic regions; dotted arrows trace data lineage.}
\label{fig:case-study}
\end{figure}
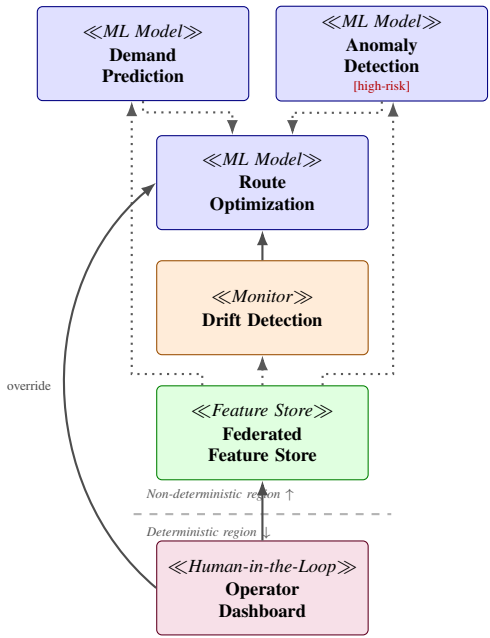

Fig.~\ref{fig:case-study} illustrates the ecosystem's C4 Component diagram with RAD-AI stereotypes, visually distinguishing AI components, non-determinism boundaries, and cross-operator data flows.

\section{Discussion}
\label{sec:discussion}

\textbf{Evaluation synthesis.} The strongest finding is the structural consistency of documentation gaps: compliance assessment and comparative analysis independently converge on the same deficiency pattern. Both Uber and Netflix exhibit identical gaps (0~of~10 concerns fully captured) despite different domains, suggesting structural rather than incidental deficiencies. The case study reveals ecosystem-level concerns invisible in standard documentation.
Together, these provide converging analytical evidence. The compliance assessment was conducted by six independent practitioners, yielding substantial inter-rater agreement ($\kappa \approx 0.68$) and mitigating author bias.

\textbf{Practical implications.} RAD-AI's backward compatibility enables incremental adoption. We suggest three stages: (1)~E1 (AI Boundary Delineation) and E2 (Model Registry View) for component visibility; (2)~E5 (AI-ADR) and E6 (AI Quality Scenarios) for decision rationale and quality commitments; (3)~E8 (Operational AI View) and E7 (AI Debt Register) for operational maturity. Organizations facing the Annex~IV deadline can prioritize extensions mapped to their risk classification using Table~\ref{tab:compliance-mapping}. The C4 AI stereotypes bridge communication between data scientists and software architects~\cite{amershi2019} by making AI components visually distinct in shared diagrams.

\textbf{Scope and next steps.} This paper establishes RAD-AI's design rationale and analytical foundations. What remains is larger-scale empirical validation: we are designing a controlled study with 10--15 software architects documenting real AI systems, measuring completeness, time-to-document, and perceived usefulness, scoped for a journal version. Complementary directions include Structurizr DSL tooling extensions and automated compliance checking. RAD-AI's current extensions target classical ML pipelines; extending them to LLM and foundation model concerns (prompt versioning, RAG pipelines, guardrails) is a priority for future work.

\textbf{Threats to validity.} \textit{Construct:} The ten Annex~IV categories and three-point scale reflect our interpretation; alternative decompositions could yield different scores. \textit{Internal:} The gap analysis (G1--G5) informed both design and evaluation, creating circularity inherent to DSR. 
The comparative analysis on independently documented production systems partially mitigates this, and our practitioner evaluation with six architects ($\kappa \approx 0.68$) further reduces author bias. Nevertheless, the small sample size and limited domains call for broader replication.
\textit{External:} Two domains (marketplace, content recommendation) and one synthetic ecosystem limit generalizability. The comparative analysis relies on public documentation that may not reflect complete internal practices. \textit{Regulatory:} Harmonized standards (prEN~18286~\cite{pren18286}) under development may introduce additional requirements.

\section{Conclusion}
\label{sec:conclusion}

Architecture documentation frameworks designed for deterministic systems are inadequate for AI-augmented ecosystems. Our comparative analysis on two unrelated production domains demonstrates that these deficiencies are structural, with identical gap patterns across both systems. RAD-AI addresses this through backward-compatible extensions (eight arc42 sections, three C4 diagram types) with a systematic EU AI Act Annex~IV compliance mapping.
Preliminary evaluation with six software architects provides evidence that RAD‑AI increases Annex IV addressability from approximately 36\% to ~93\%, captures eight AI‑specific concerns missed by standard frameworks, and surfaces ecosystem‑level needs invisible in current practice. While inter‑rater reliability indicates substantial agreement, larger empirical studies remain necessary.
By extending widely adopted frameworks rather than proposing new ones, RAD-AI enables incremental adoption while preserving existing documentation investments.


\balance


\begin{thebibliography}{45}

\bibitem{starke2023}
G.~Starke, ``arc42: The pragmatic architecture documentation template,'' 2023. [Online]. Available: \url{https://arc42.org/}. Accessed: Feb.~19, 2026.

\bibitem{brown2018}
S.~Brown, \textit{The C4 Model for Visualising Software Architecture}. Leanpub, 2018.

\bibitem{iso42010}
ISO/IEC/IEEE, ``42010:2022, Software, systems and enterprise -- Architecture description,'' International Standard, 2022. [Online]. Available: \url{https://www.iso.org/standard/74393.html}. Accessed: Feb.~19, 2026.

\bibitem{clements2010}
P.~Clements, F.~Bachmann, L.~Bass, D.~Garlan, J.~Ivers, R.~Little, P.~Merson, R.~Nord, and J.~Stafford, \textit{Documenting Software Architectures: Views and Beyond}, 2nd~ed.\hspace{0.5em} Addison-Wesley, 2010.

\bibitem{bass2021}
L.~Bass, P.~Clements, and R.~Kazman, \textit{Software Architecture in Practice}, 4th~ed.\hspace{0.5em} Addison-Wesley, 2021.

\bibitem{sculley2015}
D.~Sculley \textit{et~al.}, ``Hidden technical debt in machine learning systems,'' in \textit{Proc. NeurIPS}, 2015, pp.~2503--2511.

\bibitem{amershi2019}
S.~Amershi \textit{et~al.}, ``Software engineering for machine learning: A case study,'' in \textit{Proc. ICSE-SEIP}, 2019, pp.~291--300.

\bibitem{bosch2021}
J.~Bosch, H.~H.~Olsson, and I.~Crnkovic, ``Engineering AI systems: A research agenda,'' in \textit{Artificial Intelligence Paradigms for Smart Cyber-Physical Systems}, A.~K.~Luhach and A.~Elci, Eds.\hspace{0.5em} IGI Global, 2021, pp.~1--19.

\bibitem{nazir2023}
R.~Nazir, A.~Bucaioni, and P.~Pelliccione, ``Architecting ML-enabled systems: Challenges, best practices, and design decisions,'' \textit{J. Syst. Softw.}, vol.~207, Art.~no.~111860, 2024.

\bibitem{moin2023}
A.~Moin, A.~Badii, S.~G\"unnemann, and M.~Challenger, ``Enhancing architecture frameworks by including modern stakeholders and their views/viewpoints,'' in \textit{Proc. ICICT}, 2025, pp.~92--100.

\bibitem{struber2025}
V.~Indykov, D.~Str\"uber, and R.~Wohlrab, ``Architectural tactics to achieve quality attributes of machine-learning-enabled systems: A systematic literature review,'' \textit{J. Syst. Softw.}, vol.~223, Art.~no.~112373, 2025.

\bibitem{najafabadi2024}
F.~Amou~Najafabadi, J.~Bogner, I.~Gerostathopoulos, and P.~Lago, ``An analysis of MLOps architectures: A systematic mapping study,'' in \textit{Proc. ECSA}, ser. LNCS, vol.~14889.\hspace{0.5em} Springer, 2024, pp.~69--85.

\bibitem{mitchell2019}
M.~Mitchell \textit{et~al.}, ``Model cards for model reporting,'' in \textit{Proc. FAccT}, 2019, pp.~220--229.

\bibitem{google2022}
Google, ``Data Cards Playbook,'' 2022. [Online]. Available: \url{https://sites.research.google/datacardsplaybook/}. Accessed: Feb.~19, 2026.

\bibitem{lu2023responsible}
Q.~Lu \textit{et~al.}, ``Responsible AI pattern catalogue,'' \textit{ACM Comput. Surv.}, vol.~56, no.~7, Art.~no.~173, 2024.

\bibitem{autili2025}
M.~Autili, M.~De~Sanctis, P.~Inverardi, M.~A.~Memon, P.~Pelliccione, and S.~Pettinari, ``A reference architecture for ethical-aware autonomous systems,'' \textit{J. Syst. Softw.}, vol.~235, Art.~no.~112749, 2026.

\bibitem{euaiact}
European Parliament and Council, ``Regulation (EU) 2024/1689 laying down harmonised rules on artificial intelligence (AI Act),'' \textit{Official J. EU}, 2024. [Online]. Available: \url{https://eur-lex.europa.eu/eli/reg/2024/1689/oj}. Accessed: Feb.~19, 2026.

\bibitem{iso25059}
ISO/IEC, ``25059:2023, Software engineering -- Systems and software Quality Requirements and Evaluation (SQuaRE) -- Quality model for AI systems,'' 2023. [Online]. Available: \url{https://www.iso.org/standard/80655.html}. Accessed: Feb.~19, 2026.

\bibitem{iso42001}
ISO/IEC, ``42001:2023, Information technology -- Artificial intelligence -- Management system,'' 2023. [Online]. Available: \url{https://www.iso.org/standard/81230.html}. Accessed: Feb.~19, 2026.

\bibitem{nist2023}
NIST, ``Artificial Intelligence Risk Management Framework (AI RMF 1.0),'' 2023. [Online]. Available: \url{https://www.nist.gov/publications/artificial-intelligence-risk-management-framework-ai-rmf-10}. Accessed: Feb.~19, 2026.

\bibitem{pren18286}
CEN-CENELEC JTC~21, ``prEN~18286: Artificial intelligence -- Quality management system for EU AI Act regulatory purposes,'' 2025. [Online]. Available: \url{https://standards.iteh.ai/catalog/standards/cen/34ea911c-a980-4433-85ac-1344f93da01b/pren-18286}. Accessed: Feb.~19, 2026.

\bibitem{sovrano2025}
F.~Sovrano, E.~Hine, S.~Anzolut, and A.~Bacchelli, ``Simplifying software compliance: AI technologies in drafting technical documentation for the AI Act,'' \textit{Empir. Softw. Eng.}, vol.~30, no.~4, Art.~91, 2025.

\bibitem{aicards2024}
D.~Golpayegani, I.~Hupont, C.~Panigutti, H.~J.~Pandit, S.~Schade, D.~O'Sullivan, and D.~Lewis, ``AI Cards: Towards an applied framework for machine-readable AI and risk documentation inspired by the EU AI Act,'' in \textit{Proc. 12th Annu. Privacy Forum (APF)}, ser. LNCS, vol.~14831.\hspace{0.5em} Springer, 2024, pp.~48--72.

\bibitem{techops2025}
L.~Lucaj, A.~Loosley, H.~Jonsson, U.~Gasser, and P.~van~der~Smagt, ``TechOps: Technical documentation templates for the AI Act,'' in \textit{Proc. AAAI/ACM Conf. AI Ethics Soc. (AIES)}, 2025, pp.~1647--1660.

\bibitem{aws2025}
AWS, ``Well-Architected Machine Learning Lens,'' 2025. [Online]. Available: \url{https://docs.aws.amazon.com/wellarchitected/latest/machine-learning-lens/machine-learning-lens.html}. Accessed: Feb.~19, 2026.

\bibitem{vfunction}
vFunction, ``2025 Architecture in Software Development Report,'' 2025. [Online]. Available: \url{https://vfunction.com/resources/report-2025-architecture-in-software-development/}. Accessed: Feb.~19, 2026.

\bibitem{hermann2017}
J.~Hermann and M.~Del~Balso, ``Meet Michelangelo: Uber's machine learning platform,'' Uber Engineering Blog, 2017. [Online]. Available: \url{https://www.uber.com/blog/michelangelo-machine-learning-platform/}. Accessed: Feb.~19, 2026.

\bibitem{sun2020}
C.~Sun, N.~Azari, and C.~Turakhia, ``Gallery: A machine learning model management system at Uber,'' in \textit{Proc. EDBT}, 2020, pp.~474--485.

\bibitem{uberd3}
Uber Engineering, ``D3: An automated system to detect data drifts,'' Uber Blog, 2023. [Online]. Available: \url{https://www.uber.com/blog/d3-an-automated-system-to-detect-data-drifts/}. Accessed: Feb.~19, 2026.

\bibitem{uber2024}
Uber Engineering, ``From predictive to generative: How Michelangelo accelerates Uber's AI journey,'' Uber Engineering Blog, 2024. [Online]. Available: \url{https://www.uber.com/blog/from-predictive-to-generative-ai/}. Accessed: Feb.~19, 2026.

\bibitem{netflix2024}
Netflix Technology Blog, ``Supporting diverse ML systems at Netflix,'' 2024. [Online]. Available: \url{https://netflixtechblog.com/supporting-diverse-ml-systems-at-netflix-2d2e6b6d205d}. Accessed: Feb.~19, 2026.

\bibitem{netflixflink2024}
M.~Cho and M.~Liu, ``Building a scalable Flink platform: A tale of 15,000 jobs at Netflix,'' in \textit{Confluent Current}, 2024. [Online]. Available: \url{https://current.confluent.io/2024-sessions/building-a-scalable-flink-platform-a-tale-of-15-000-jobs-at-netflix}. Accessed: Feb.~19, 2026.

\bibitem{kreuzberger2023}
D.~Kreuzberger, N.~K\"uhl, and S.~Hirschl, ``Machine learning operations (MLOps): Overview, definition, and architecture,'' \textit{IEEE Access}, vol.~11, pp.~31\,866--31\,879, 2023.

\bibitem{wieringa2014}
R.~Wieringa, \textit{Design Science Methodology for Information Systems and Software Engineering}.\hspace{0.5em} Springer, 2014.


\bibitem{stol2018}
K.-J.~Stol and B.~Fitzgerald, ``The ABC of software engineering research,'' \textit{ACM Trans. Softw. Eng. Methodol.}, vol.~27, no.~3, Art.~no.~11, 2018.

\bibitem{madr}
``MADR: Markdown Any Decision Records,'' 2024. [Online]. Available: \url{https://adr.github.io/madr/}. Accessed: Feb.~19, 2026.

\bibitem{bogner2021}
J.~Bogner, R.~Verdecchia, and I.~Gerostathopoulos, ``Characterizing technical debt and antipatterns in AI-based systems: A systematic mapping study,'' in \textit{Proc. TechDebt}, 2021, pp.~64--73.

\bibitem{sidhpurwala2025}
H.~Sidhpurwala, E.~Fox, G.~Mollett, F.~Cano~Gabarda, and R.~Zhukov, ``Blueprints of trust: AI system cards for end-to-end transparency and governance,'' \textit{arXiv:2509.20394}, 2025.

\bibitem{modelcardpp}
M.~Boone, N.~Pope, D.~Yared, C.~Xiao, and A.~Anandkumar, ``Enhancing AI transparency and ethical considerations with Model Card++,'' NVIDIA Technical Blog, 2022. [Online]. Available: \url{https://developer.nvidia.com/blog/enhancing-ai-transparency-and-ethical-considerations-with-model-card/}. Accessed: Feb.~19, 2026.

\bibitem{metasystemcards}
Meta~AI, ``System cards, a new resource for understanding how AI systems work,'' Meta AI Blog, 2022. [Online]. Available: \url{https://ai.meta.com/blog/system-cards-a-new-resource-for-understanding-how-ai-systems-work/}. Accessed: Feb.~19, 2026.

\bibitem{bucaioni2025}
A.~Bucaioni, M.~Weyssow, J.~He, Y.~Lyu, and D.~Lo, ``Artificial intelligence for software architecture: Literature review and the road ahead,'' \textit{arXiv:2504.04334}, 2025.

\bibitem{esposito2026}
M.~Esposito, X.~Li, S.~Moreschini, N.~Ahmad, T.~Cerny, K.~Vaidhyanathan, V.~Lenarduzzi, and D.~Taibi, ``Generative {AI} for software architecture. {Applications}, challenges, and future directions,'' \textit{J. Syst. Softw.}, vol.~231, Art.~no.~112607, 2026.

\bibitem{bucaioni2025aac}
A.~Bucaioni, A.~{Di Salle}, L.~Iovino, P.~Pelliccione, and F.~Raimondi, ``Architecture as code,'' in \textit{Proc. 22nd IEEE Int. Conf. Softw. Archit. (ICSA)}, Odense, Denmark, 2025, pp.~187--198.

\bibitem{bucaioni2025ra}
A.~Bucaioni, M.~Weyssow, J.~He, Y.~Lyu, and D.~Lo, ``A functional software reference architecture for LLM-integrated systems,'' in \textit{Proc. 22nd IEEE Int. Conf. Softw. Archit. Companion (ICSA-C)}, Odense, Denmark, 2025, pp.~1--5.

\bibitem{jarvenpaa2024}
H.~J{\"a}rvenp{\"a}{\"a}, P.~Lago, J.~Bogner, G.~Lewis, H.~Muccini, and I.~Ozkaya, ``A synthesis of green architectural tactics for {ML}-enabled systems,'' in \textit{Proc. 46th IEEE/ACM Int. Conf. Softw. Eng.: Softw. Eng. Soc. (ICSE-SEIS)}, Lisbon, Portugal, 2024, pp.~130--141.

\end{thebibliography}
\end{document}